# A MINIMAL MODEL FOR QUANTUM GRAVITY

**Paola A. Zizzi**

*Dipartimento di Matematica Pura ed Applicata, Università di Padova,*
*Via Belzoni, 7, 35131 Padova,*
*Italy*
zizzi@math.unipd.it

**Abstract**

We argue that the model of a quantum computer with N qubits on a quantum space background, which is a fuzzy sphere with $n = 2^N$ elementary cells, can be viewed as the minimal model for Quantum Gravity. In fact, it is discrete, has no free parameters, is Lorentz-invariant, naturally realizes the Holographic Principle, and defines a subset of punctures of spin networks' edges of Loop Quantum Gravity labelled by spins $j = 2^{N-1} - \frac{1}{2}$. In this model, the discrete area spectrum of the cells, which is not equally spaced, is given in units of the minimal area of Loop Quantum Gravity (for j=1/2), and provides a discrete emission spectrum for quantum black holes. When the black hole emits one string of N bits encoded in one of the n cells, its horizon area decreases of an amount equal to the area of one cell.

**Key words:** Quantum Computers, Non-commutative Geometry, Loop Quantum Gravity, Holographic Principle, Quantum Black Holes.

## 1. Introduction

The two main roads to quantum gravity are String Theory and Loop Quantum Gravity (LQG). The latter is background independent, and will be the one taken into account in the context of this paper.

The main results of LQG (see [1] for recent reviews) are the discreteness of geometry at the Planck scale, and the computation of black holes entropy.

However, LQG is plagued by the appearance of a free parameter, called the Immirzi parameter [2], which arises from an ambiguity in the quantization formalism, and by other problems, like the Hamiltonian constraint.

We believe that, to try to fix such problems, one should also build a minimal theory of Quantum Gravity, without starting from quantization of the classical theory, and then compare the results. Then, we look for a quantum system, on a quantum (non-commutative) space background, (for reviews on quantum spaces, see [3]), which could mimic (simulate) space-time at the Planck scale. The first part of this theoretical construction was done in a recent paper [4], where we found a model for a quantum computer (for an introduction to quantum computing, see [5]) on a quantum space, namely the fuzzy sphere [6]. There, we used two powerful mathematical tools: the Gelfand-Naimark-Segal construction [7], and the non-commutative version of the Gelfand-Naimark theorem [8]. In addition, we developed a logical interpretation of the reversible measurement, which is allowed in such a model [9].

## 2. Computational Loop Quantum Gravity

In the present paper, we will maintain the theoretical construction developed in [4] and will accomplish the quantum gravity model by exploiting the Holographic Principle [10]. The resulting model is that of a quantum computer (the quantum system) on a fuzzy sphere (the quantum space) which turns out to be a naturally holographic model.

The quantum computer on the quantum space provides a "simulation" of quantum gravity via the holographic map, which is an isomorphism. The resulting simulation is what we call "minimal model for quantum gravity". The isomorphism between the quantum system and the quantum space is in accordance with a background independent theory of quantum gravity, like Loop Quantum Gravity. Actually we give a model for "Computational Loop Quantum Gravity" (CLQG).

In summary, this minimal model for QG is a consequence of the isomorphism between the quantum computer, and the fuzzy sphere, and of the holographic structure arising from it.

In [4], we used the C*-formulation of QM, by which the basic structure of a quantum system is the C*-algebra [11] of linear bounded operators. In this formulation, a state is a complex positive linear functional. From a state, one can then build a unitary representation by using the GNS construction. It should be noticed that if the state is pure, the resulting representation is irreducible. The advantage of the C*-formulation is that to any non-commutative C*-algebra is associated, by the GN theorem, a quantum space.

In our case, the non-commutative C*-algebra is the algebra of logic quantum gates, which are $n \times n$ unitary matrices, where $n = 2^N$, and N is the number of qubits. By the GNS construction, from the $n$ pure states of this algebra, which correspond to $n$ cyclic vectors of the associated Hilbert space $C^{2N}$, one can build all the $j = 2^{N-1} - \frac{1}{2}$ irreducible representations of SU(2). By the GN theorem, to this non-commutative C*-algebra, it is associated a quantum space which is the fuzzy sphere with $n = 2^N$ cells.

The $n$ cells are the $n$ pure states, and then correspond to the $n$ cyclic vectors, which are strings of N bits. In this paper, the isomorphism between cells and bit strings arising from the GNS construction and the GN theorem, will be interpreted as a natural realization of the Holographic Principle. In fact, the elementary cells do not simply encode bits of information, as in the Holographic Principle, the cells *are* bit strings, because of the above mentioned isomorphism. It should be noticed that this model is quite adequate for being a simulation of quantum gravity.

It is generally believed that space-time at the Planck scale is discrete. However, a discrete space like a lattice breaks Lorentz invariance. One of the simplest ways [12] to introduce discreteness at the Planck scale in a Lorentz-invariant way is by considering non-commutative space-time coordinates. An example of discrete, and still, Lorentz invariant space is just the fuzzy sphere. In fact, the fuzzy sphere is discrete (points are replaced by elementary cells) and, obviously, the fuzzy sphere, having the same topology of an ordinary sphere $S^2$, is invariant under SO(3) rotations.

The question is now whether the resulting model has indeed any link to Loop Quantum Gravity, and to current models of quantum black holes, and the answer is in the affirmative. In fact, as we will see in more detail in what follows, the number N of qubits is equal to the number of punctures of spin networks' edges in the $j = 2^{N-1} - \frac{1}{2}$ irreducible representations of SU(2).

This model is endowed with two discrete area spectra, one for the fuzzy sphere, and one for the cells. Both of them turn out to be expressed in multipliers of the minimal area of Loop Quantum gravity, that is the area created by one puncture of a spin network's edge in the j=1/2 representation of SU(2).

The area spectrum of the fuzzy sphere, which is equally spaced, provides the same black hole emission spectrum as that of Bekenstein and Mukhanov [13]. However, in our context, this area spectrum is unphysical, as it corresponds to the emission of one qubit in a superposed state. This cannot happen, as absorption and emission of information by a black hole correspond, at least in our model, to the classical input and output, respectively, in quantum computing. The superposed state can exist only on the surface of the fuzzy sphere, that is the black hole surface horizon (the boundary), corresponding to the quantum computational process in the bulk (the quantum computer).

The cellular area spectrum, which is not equally spaced, has instead a physical meaning. When the black hole emits one string of N bits encoded in one of the n cells, the decrease of the horizon area is equal to the area of one cell. In other words, as the superposed state of $2^N$ qubits can be accommodated in the $n = 2^N$ cells of the fuzzy sphere, taking out one cell, is equivalent to make a quantum measurement, (or entanglement with the environment), breaking the superposition, and getting a classical output of N bits.

In what follows, we will give some more technical details about the model under study.

The fuzzy sphere is constructed [6] replacing the algebra of polynomials on the sphere $S^2$ by the non commutative algebra of complex $n \times n$ matrices, which is obtained by quantizing the coordinates $x_i$ (i=1,2,3): $x_i \to X_i = kJ_i$, where the $J_i$ form the n-dimensional irreducible representation of the algebra of SU(2): $[J_i, J_j] = i\varepsilon_{ijk}J^k$, and k is the non-commutativity parameter [6]:

$$k = \frac{r}{\sqrt{n^2 - 1}} \qquad (1)$$

Where r is the radius of the $S^2$ sphere, and $n = 2j + 1$ is the number of elementary cells.

It should be noticed that, in the quantization process, the spherical harmonics on $S^2$, are mapped [14] into $n \times n$ matrices whose components are given by 3-j symbols, corresponding to vertices of spin networks.

The area of an elementary cell is:

$$A_{EC} = \frac{4\pi r^2}{\sqrt{n^2 - 1}} \qquad (2)$$

And the total area of the fuzzy sphere is:

$$A_{FS} = n \frac{4\pi r^2}{\sqrt{n^2 - 1}} \qquad (3)$$

For large n: $A_{FS} \to A_{S^2}$

Recall that, in our case, $n = 2^N$, where N is the number of qubits, then (2) and (3) can be rewritten, respectively as:

$$A_{EC} = \frac{4\pi r^2}{\sqrt{(2^N - 1)(2^N + 1)}} \qquad (4)$$

$$A_{FS} = 2^N \frac{4\pi r^2}{\sqrt{(2^N - 1)(2^N + 1)}} \qquad (5)$$

The Holographic Principle [10] states that, at the quantum level, the state of a physical system located in a region of space of volume V, is fully characterized by the degrees of freedom on the boundary surface of V. The number of degrees of freedom is then related to the area A of the boundary surface, instead of the volume V. More precisely, each pixel of area (a Planck area $l_P^2$, where $l_P \approx 10^{-33} cm$, is the Planck length) encodes one bit of information.

The number of physical degrees of freedom (the entropy of the system in V) is less or equal to one fourth of the area A of the surface measured in Planck-area units: $S \leq \frac{A}{4 l_P^2}$ the equality being fulfilled by black holes. Then, the Holographic principle can be considered as a generalization of the Bekenstein-Hawking entropy bound [15].

In our model, due to the isomorphism arising from the GNS construction, the $2^N$ degrees of freedom of the bulk (the quantum computer) are the $2^N$ elementary cells of the boundary surface (the fuzzy sphere), every cell encoding one string of N bits. Then, our model naturally realizes the Holographic Principle, discussed above, although in a generalized way.

Consequently, in the case of fuzzy black hole-quantum computers, the Bekenstein-Hawking entropy bound is saturated:

$$S = \frac{A_{FS}}{4 l_P^2} \qquad (6)$$

Where: $l_P \approx 10^{-33} cm$, is the Planck length, and S is the entropy, which in our case, is just the information entropy of N qubits:

$$S = N \ln 2 \qquad (7)$$

By (5), (6) and (7), one gets:

$$A_{FS} = N A_0 \qquad (8)$$

And consequently:

$$A_{EC} = \frac{N}{2^N} A_0 \qquad (9)$$

Where:

$$A_0 = 4 \ln 2 \, l_P^2 \qquad (10)$$

In Loop Quantum Gravity, the discrete area spectrum [16] is:

$$A = 8\pi l_P^2 \gamma \sum_i \sqrt{j_i (j_i + 1)} \qquad (11)$$

Where the $j_i$ are the spins in the irreducible representation of SU (2), which label the spin networks' edges, and $\gamma$ is the Immirzi parameter.

Then, $A_0$ in (10) is the minimal area of LQG, that is the area created by one puncture of a spin network's edge in the smallest representation of SU(2), that is, j=1/2, for the value of the Immirzi parameter:

$$\gamma = \gamma_0 = \frac{\ln 2}{\pi\sqrt{3}} \qquad (12)$$

This value of the Immirzi parameter, which was obtained [17] in the computation of black holes entropy, in the case of all equal punctures in the j=1/2 representation of SU (2), is then confirmed [18] by our minimal model which has no free parameters on its own.

The two discrete spectra in (8) and (9) can be interpreted as follows.

The discrete spectrum in (8) is relative to the area levels of the fuzzy sphere with a quantum computer of N qubits working upon. By comparing this spectrum with that of Loop Quantum Gravity, with all punctures of spins j=1/2:

$$A_{1/2} = 4\ln 2\, l_P^2 \nu \qquad (13)$$

where $\nu$ is the number of punctures, we see that the number N of qubits encoded in the fuzzy sphere is equal to the number $\nu$ of punctures on the sphere $S^2$.

However, from the spectrum (9) we see that, to each puncture in LQG, correspond $2^\nu$ elementary cells in our model. From (8), (9) and (13) we get the following scheme.

N=1 qubits $\leftrightarrow$ $\nu$ =1 punctures with j=1/2

$A_{FS(1)} = A_0$

n=2 cells

$A_{EC(1)} = \frac{1}{2} A_0$ .

Each of the two cells encodes one bit of information: $|0\rangle$ and $|1\rangle$

N=2 qubits $\leftrightarrow$ $\nu$ =2 punctures with j=3/2

$A_{FS(2)} = 2A_0$

n=4 cells

$A_{EC(2)} = \frac{1}{2} A_0 = A_{EC(1)}$

Each of the four cells encodes a string of 2 bits: $|00\rangle, |01\rangle, |10\rangle, |11\rangle$

N=3 qubits $\leftrightarrow$ $\nu$ =3 punctures with j=7/2

$A_{FS(3)} = 3A_0$

n=8 cells

$A_{EC(3)} = \frac{3}{8} A_0$

Each of the eight cells encodes a string of 3 bits: $|000\rangle, |001\rangle, |010\rangle, |011\rangle,$
$|100\rangle, |101\rangle, |110\rangle, |111\rangle$.

And so on. The case $N = \nu = 1$ is illustrated for both LQG and for our model in Fig.1.
The case $N = \nu = 2$ is illustrated in Fig. 2 for LQG, and in Fig. 3 for our model.

**3. The discrete emission spectrum**

The Bekenstein bound is saturated for black holes. We have tacitly assumed, by (6), that a quantum black hole is a quantum computer on a fuzzy sphere.

The modelling of a black hole like a fuzzy sphere was originally conceived by 'tHooft [19], although in that case there was not any relation to a quantum computer.

In our model, when the black hole emits a string of N bits, the area of the surface horizon decreases of the area of one of the $n = 2^N$ cells, that is the decrease of area is just given by the area of one cell in (9)

$$\Delta A_N = A_{EC(N)} = \frac{N}{2^N} A_0 \qquad (14)$$

From the relation between the area A and the mass M of a black hole: $A = 16\pi G^2 M^2$
Where G is the gravitational Newton constant, we get a decrease of the mass:

$$\Delta M_N = \frac{N}{2^N} \frac{\ln 2 \hbar}{8\pi GM} \qquad (15)$$

And an emission frequency spectrum:

$$\omega_N = \frac{N}{2^N} \omega_{BM} \qquad (16)$$

Where:

$$\omega_{BM} = \frac{\ln 2}{8\pi GM} \qquad (17)$$

Is the Bekenstein-Mukhanov fundamental frequency [13], which is of the same order of Hawking maximum frequency, $\omega_H$ [20]. A similar argument is the one given by Hod [21] and later by Dreyer [22], based on the spectrum of quasi-normal modes derived from classical relativity.
The discrete area spectrum for the fuzzy sphere, in (8) gives the discrete equally spaced frequency spectrum:

$$\omega_N' = N\omega_{BM} \qquad (18)$$

That is, the same emission spectrum of Bekenstein-Mukhanow [13].
However, as we said above, the spectrum in (18) in our case is unphysical, as it would correspond to the emission of one qubit in a superposed state.
The reason is the following. In the Bekenstein-Mukhanov case, one elementary cell just corresponds, in Loop Quantum Gravity, to one puncture, and the cell, which has area $A_0$, encodes one bit, either $|0\rangle$ or $|1\rangle$, not a superposed state of them, as in our case. Recall that the cellular area encoding one bit is, in our case: $\frac{1}{2} A_0$.

Instead, we take the frequency spectrum (16) as the physical one, as it corresponds to the more realistic case of the emission of one string of N bits. For example, in the simple case N=1, either $|0\rangle$ or $|1\rangle$ is emitted, breaking the superposed state, and one of the two elementary cells evaporates, leaving one cell of area equal to $\frac{1}{2} A_0$, which still encodes the other bit. See Fig. 4.

In this simple model, the remnant of a black hole would have a surface horizon with an area that is one half the fundamental area of Loop Quantum Gravity. However, it would not be a mini-quantum computer in the sense that it will encode a classical bit, not a qubit.
It should be noticed that, in our case, available (classical) information is emitted, in terms of bit strings, by the evaporating black hole. Thus, in the present model, there is no information loss, differently from earlier Hawking's assumption [23].
To conclude, we would like to stress again that our results are in accordance with LQG, for the subset of $\nu = \ln_2(2j+1)$ punctures with spins j=1/2, 3/2, 7/2….
However, the information content is still different: while in our case, each elementary cell encodes a string of N bits, in LQG, each elementary cell encodes one bit. Nevertheless, this difference is not too surprising. In fact, in LQG, the surface horizon of a Schwartzschild black hole, with a classical SO (3) symmetry, is a classical (punctured) sphere. The latter is not a quantum space in the sense of non-commutative geometry, thus cannot be put in a one-to-one correspondence with a quantum system like a quantum computer by the GNS construction and the GN theorem.

**Acknowledgements**
Work supported by the research project "Logical Tools for Quantum Information Theory", Department of Pure and Applied Mathematics, University of Padova. I am grateful to C. Rovelli and M. Matone for discussions and advices.

**Fig.1**

**The case N=1 for both LQG and MM**

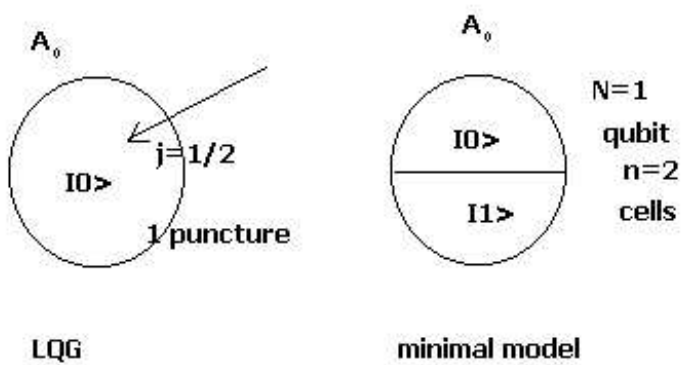

LQG                minimal model

**Fig. 2**

**The case N=2 for LQG: 2 punctures**

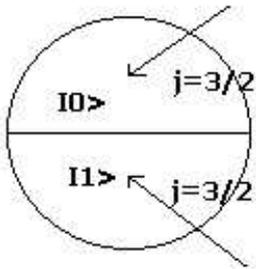

**Fig. 3**

**The case N=2 for the MM: two qubits, four cells**

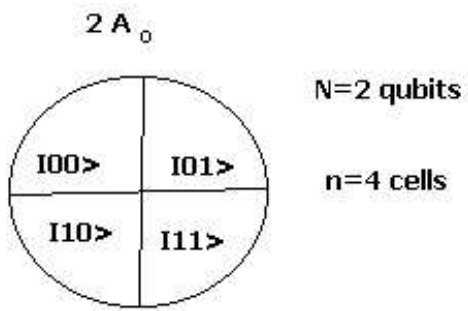

**Fig. 4**

**Emission of one bit from a black hole**

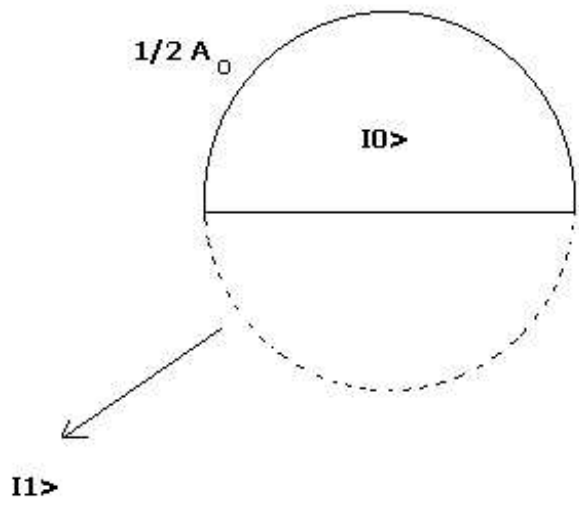